# Simulation based Study of TCP Variants in Hybrid Network

Wafa Elmannai, Abdul Razaque and Khaled Elleithy
welmanna@bridgeport.edu    arazaque@bridgeport.edu    elleithy@bridgeport.edu

**Department of Computer science and Engineering**

**University of Bridgeport, Bridgeport, CT 06604**

**Abstract:**
Transmission control protocol (TCP) was originally designed for fixed networks to provide the reliability of the data delivery. The improvement of TCP performance was also achieved with different types of networks with introduction of new TCP variants.

However, there are still many factors that affect performance of TCP. Mobility is one of the major affects on TCP performance in wireless networks and MANET (Mobile Ad Hoc Network).

To determine the best TCP variant from mobility point of view, we simulate some TCP variants in real life scenario. This paper addresses the performance of TCP variants such as TCP-Tahoe, TCP-Reno, TCP-New Reno, TCP-Vegas, TCP-SACK and TCP-Westwood from mobility point of view.

The scenarios presented in this paper are supported by Zone routing Protocol (ZRP) with integration of random waypoint mobility model in MANET area. The scenario shows the speed of walking person to a vehicle and suited particularly for mountainous and deserted areas. On the basis of simulation, we analyze Round trip time (RTT) fairness, End-to-End delay, control overhead, number of broken links during the delivery of data. Finally analyzed parameters help to find out the best TCP variant.

## I. INTRODUCTION

The fact that TCP is originally started with the wireless network, even its well performance in wired network. [16] As well as it is cleared that the deployment of wireless networks in the past few years have motivated lot of people to study and make efforts for improving the performance of TCP in wireless networks and all this work confirmed that TCP in its present structure is not appropriate for MANET. Since the TCP has over head of packet loss which is may cause the buffer congestion because the MANETs losses is due to error or the frequently of the mobility .So, the TCP can't be efficient probably [15].

Furthermore, the main reason of the weak performance of the standard TCP in the wireless network is the inability of discovers the packet loss. So, we can realize that both characteristics have same reason of packet loss problem which is basic on the network congestion. [16]

However, some new protocols have been proposed and implemented. We will evaluate some of these protocols, and we will demonstrate how they increase the performance on wireless networks. This study will be based on Hybrid network particularly MANET.

By going back to the MANET which is also has increased due to the spreading of inexpensive portable and computing devices. The MANET network is special network due to the ability of its nodes to communicate with each other through packed forwardly by the intermediate nodes. So, it can be set up in any remote areas without infrastructure support. The nodes, which are part of MANET and they require data and information from database but database is available in the wired network, therefore MANET can be integrated with wired network to obtain the[14] required data and information. Some applications run over the database and these applications are supported by Transmission Control protocols. This paper aims to exhibit the flaws for TCP compare to Hybrid network especially MANET. Exhibition will be done with ns2. This simulator will provide the outcomes for different protocols' throughput, broken links overhead etc…that we will analyze. TCP uses some congestion control parameters, which include congestion window, recovery mechanism, retry limit, maximum packet size and back up mechanism for IEEE 802.11 retransmission [2, 5]. To minimize the congestion problems, as different TCP Variants have been introduced & simulated on various schemes in order to identify the performance for each TCP Variants and analyzing which variant has considerable performance due to mobility. To determine the performance for each TCP Variants, as new architecture and approaches are required to find out complete behavior of the variants. This motivation results to introduce such network to analyze the effectiveness for each TCP Variants. The random waypoint mobility model is incorporated to control the moments of nodes. In order to analyze the impact of mobility while simulating the Hybrid network, it is essential that underlying mobility model attain realistic scenario or at least important feature. To this conclusion, we deem that this paper makes reasonable contribution. The rest of the paper is organized as follows. In Section II, We describe problem statement. In Section III, we present related work. In Section IV, we present Design of Hybrid network Scenario. In Section V, we define Setup of Initial connection and Hand off process. In Section VI, We give Overview of mobility model and simulation setup. In Section VII, We present simulation Results, In Section VIII; we talk about discussion of Results. Finally Section IX concludes the work and future directions.

## II. RELATED WORK

Ramarathinams et al. [18] evaluated the performance of TCP Reno, New Reno, SACK and Tahoe with respect to Goodput under three routing protocols over static multi-hop network and claimed that Reno got better results but scenario is not fully explain. Our work is completely different from their work. We introduce ZRP protocol in Hybrid network with inclusion of additional TCP Vegas and Westwood. Our work almost discusses all the issues of Manet due to TCP Variants and routing protocol. Abdul Razaque et al. [7] compared TCP Variants in APN Hybrid network by using DSR routing protocol. We previously focused on Throughput, Packet delivery ratio and End-to-End delay but here point out the performance of some existing TCP Variants from different angles and incorporated Hybrid routing protocol "ZRP" in our architecture. We focus on control overhead, In-order delivery of data, broken links, RTT from different perspective. The finding of this paper gives complete knowledge about the behavior of each TCP Variants in Hybrid environment. We narrowly analyze all issues of Manet due to mobility and showed their affect on the performance of each TCP Variants.A.O. Oluwatope et al. [5] Used the realistic scenario of Hybrid network and simulated the TCP Reno, TCP SACK and TCP Westwood. They claimed that TCP Westwood was better performer in their static scenarios whereas our work was completely depends on mobility and speed with Random way point model. We have thoroughly studied the behavior of TCP Variants.

## III. PROBLEM STATEMENT

The scope of this paper is to analyze some existing TCP Variants over Hybrid network. The focus of study is particularly around the performance metrics such as throughput, RTT fairness, End-to-End delay, broken links and control overhead due to mobility. The major contribution of this research is to identify the loss of Goodput on different mobility ratios and to design mobility based Hybrid network with Random Waypoint Mobility model, where TCP Variants will be simulated and analyzed from mobility point of view.

## IV. DESIGN OF HYBRID NETWORK SCENARIO

We have designed random waypoint mobility aware scenario in Hybrid network by combining the features of wired network with wireless and MANET in order to make reasonable communication even in remote areas. The nodes, which make the possible communication between different segments of network are called gateway (Anchor Point Node).

The APN can play a role as coordinator in the network. Three segments of networks are jointly connected to make the Hybrid network. The APNs are located on different positions. The gateway(APN) of MANET has information about the nodes, and these nodes are assigned the IPs locally through Dynamic Host Configuration Protocol Server (DHCP). An APN that is part of MANET is said to be MANET Anchor Point Node (MAPN) similarly, the node that is located at the area where wireless range becomes weak is called Infrastructure Based Anchor Point Node (IBAPN). Both APNs can play a role as coordinators and formulate possible communication for rest of nodes in fixed and MANET segment of network. The wired and wireless segments of network cover the urban and suburbs areas of urban environment and MANET portion of network covers remote areas. We use random waypoint (RW) mobility model at different mobility ratios and speeds in this network. The MANET network is routed with ZRP. We have created twenty traffic flows to analyze the performance of TCP Variants. This Hybrid network could be suited for urban and remote areas given in figure1.

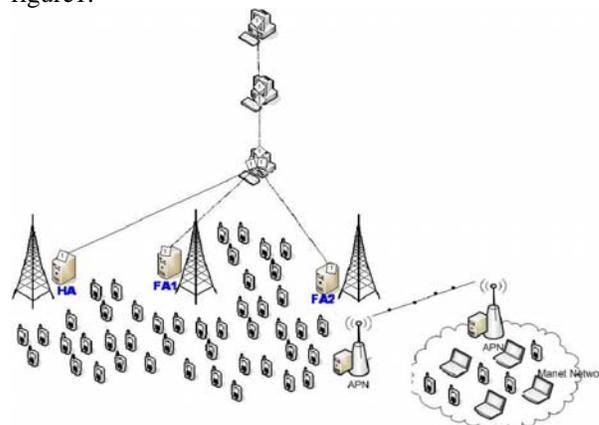

Figure 1: Design of Hybrid Network Scenario [3]

## V. SETUP OF INITIAL CONNECTION AND HAND OFF PROCESS

The section presents outline of initial connection setup and handoff process for Manet Mobile Node (MMN). Figure 2 shows timing diagram and describe the signals involved in it. Initial connection setup and MMN hand off process can be defined in the following steps. Initially the nodes, which are the part of Manet, intending to communicate with corresponding node (CN). They should establish initial connection setup and send the message through Current MANET Anchor Point Node (CMAPN) "Request for connection setup with CN".

1. When CMAPN obtains the Request for connection setup from MMN and forwards the message "coordination request for connection setup" to Infrastructure Based Anchor Point Node (IBAPN). In response CMAPN also sends back message "Reply for connection setup with CN" to (MMN). When MMN obtains the message from CMAPN then it will be waiting till initial connection is established. 2. IBAPN forwards the message "forwarding coordination request for connection setup" to respective HA/FA within wired area. HA/FA informs the IBAPN with message "Accept coordination request" to CMAPN.
3. HA/FA forwards message with "forwarding initial connection setup" to (CN) and sends back response to IBAPN "Accept forwarding coordination request". When CN receives the message then inform the HA/FA "Accept initial connection setup" with (MMN). With establishment of initial connection setup between CN and MMN then data exchange process is started.
4. When MMN changes the location and moves to other MANET then it sends the request for handoff to new MANET Anchor point node (NMAPN) with message "request for joining".
5. NMAPN sends the message "location change forwarding message" (LCFM) to IBAPN for informing the handoff process and similar message is forwarded to HA/FA and finally to CN for location update.
6. NMAPN forwards the LCFM to IBAPN and also "update" to CMAPN. In response, CMAPN sends acknowledgement (ACK) to NMAPN for location update.
7. When CN gets the message LCFM then it sets the connection again with MMN and message is forwarded with "new connection setup in change of location".
8. With the establishment of new connection, the data exchange process is initiated.

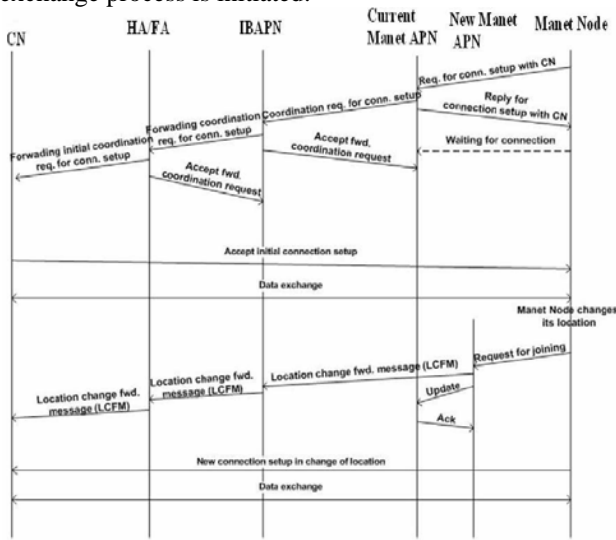

Figure 2: Initial connection setup and Hand off process [1].

## VI. OVERVIEW OF MOBILITY MODEL & SIMULATION SETUP

The purpose of this paper is to investigate the results of TCP Variants under mobility Based Hybrid Network. In this paper, we have critically evaluated the Performance of TCP Variants with respect to different mobility ratios and analyzed the behavior of each TCP Variants. The parameters of interest include throughput, RTT fairness, In-order delivery of data, effect of mobility on Goodput, End-to-End delay and control overhead.

## VI. (A) OVERVIEW OF RANDOM WAYPOINT MOBILITY MODEL

The literature survey gives the detail of mobility models in [6] but these mobility models mostly described theoretically and major variation in mobility patterns are found in real scenarios. The new mobility models were also introduced in [9], such as free Mobility models (FM), Manhattan Mobility model ((MM) and Reference Point Group Mobility model (RPGM). The social network Mobility model (SNM) is discussed [7].The research community mostly uses the Random Waypoint (RW) Mobility model in Mobile Adhoc networks. RW Mobility model is also incorporated in ns2 and detail of this model is given in following Para: The nodes move to random destination with given velocity by using normal or uniform distribution [Velocity minimum, Velocity maximum] when nodes reach the destination, they stop for the time given by the "pause" time. The pause time can be constant value or uniform distribution [0, time pause maximum]. After completion of pause time, mobile nodes decide the destination and direction randomly and this process continues till the simulation time ends.

## VI. (B) SIMULATION SETUP

Ns2.28 on Red Hat 8 is used for simulation. The random Waypoint mobile scenario is generated. The simulator gives a proper model for signal propagation and transmission range is 250 meter [8]. The sensing and interference range is 550 meter. TCP New Reno, Reno, Tahoe, SACK, Westwood and Vegas are simulated and investigated on the same network so as to ensure fairness and behavior of the TCP Variants. The length of packet is 1040 bytes including 40 bytes are overhead. In this simulation, 40 mobile nodes both in wireless and MANET segment of network are placed. As we check the mobility of MANET-nodes, which move within rectangular field of 600 *1200 meters. RW generates mobile scenario and start location of nodes. Constant values for pause time have been set, which are 10 seconds after each 50 seconds. Total simulation time is 300 seconds. The minimum speed of the node (Vmin) is 0 m/sec and maximum speed (Vmax) are 10 m/sec respectively. The moving speed of node is randomly obtained through uniform division [Vmin, Vmax]. We run simulations, which cover combination of the pause time and moving speed of nodes. The percentage of mobility means how many mobile nodes move and resulting how many links break in the MANET. Hence 50% mobility shows 20 nodes move out of 40 nodes. Zone Routing Protocol (ZRP is locally proactive and globally reactive , which gives better performance for routing in multi-hop Mobile Adhoc Network (MANET) and produces minimum routing overhead. It has also capability to deliver approximately all originated data packets, even with perpetual, rapid movement of all nodes in the network. The major cause for better performance is that ZRP functions completely on demand with no periodic motion of any type mandatory at any stage in the network.

## VII. SIMULATION RESULT
In this subsection, we discuss the results of simulated scenario.

### A. (THROUGHPUT FOR TCP VARIANTS AT DIFFERENT MOBILITY RATIOS)
We have simulated Hybrid network Scenario with network simulator-2 and analyzed the performance of TCP Variants (Reno, Vegas, New Reno, Tahoe, Westwood and SACK). We have collected acknowledged packets for each TCP Variants and analyzed the throughput performance. The figure 3 shows the throughput performance for each TCP Variants at the maximum speed of 5 m/sec with Random Waypoint Mobility model. The performance gradually decreases to each TCP Variants from 5% mobility to 50% mobility. The reasons for decreasing rate of delivered packet is mobility since 2 nodes move in the network that less links breaks and takes less time to recover whereas 20 mobility nodes cause more time to recover from broken links. If MANET segment of network is part of Hybrid network, the topology of MANET remains mostly dynamic and major affecting factors are radio channel fading and mobility of nodes [15].The mobility also degrades the performance of TCP Variants because mobility causes the change of routing information in network, which causes long RTT and repeated timeouts resulting takes long time in retaining. Due to mobility, the receiver gets out of order segments resulting in the receiver generates Acknowledgements (Ack) only for highest in-order packets. This causes the duplicate Acks and fast retransmission algorithm starts and congestion window reduces. Therefore ssthresh and cwnd are set to max (unacknowledged data/2,2_MSS) & ssthresh+3MSS.TCP Vegas delivers more packets and TCP Tahoe and Reno relatively send same amount of Packets. The reason of delivering the more packets for TCP Vegas is that TCP Vegas retransmits the lost packets after receiving the 2 duplicate acknowledgements whereas other TCP Tahoe, Reno, New Reno, SACK and Westwood retransmit the segments after receiving the three duplicate acknowledgments but in some cases third (dupack) takes either long time or does not receive third dupack and timeout expires. It is advantage of Vegas over above TCP variants because TCP Vegas mostly retransmits the lost segments before Retransmission time out (RTO). The other reason is that TCP Vegas does not wait for loss to trigger congestion window (cwnd) reduction. Vegas possesses interesting approach regarding the congestion because it estimates the level of congestion before it occurs rather try to avoid it. The level of congestion is measured on basis of sample RTT and size of sending window that is also the reason; the Sender estimates the current throughput against every RTT [11]. TCP Tahoe and Reno have delivered fewer packets than rest of TCP Variants.

TCP Tahoe faces the problem due to repeat of slow start phases on each dropped segment, particularly when error is transient and not constant. In this case congestion window shrinks and bandwidth cannot fully be utilized. Fast Recovery algorithm for TCP New Reno can degrade the performance due to multiple losses of packets during single window because Fast Recovery algorithm can manage only single loss per RTT [11, 17]. TCP Tahoe and Reno Variants have more difficulty to differentiate between loss and congestion in wireless environment such as IEEE 802.11.The performance degrading factor for TCP Reno and Tahoe is also size of congestion window because these variants cannot send data during the timeout period, if mainly packets loss occurs [14]. The findings of our simulation were also validated by researchers in different papers.

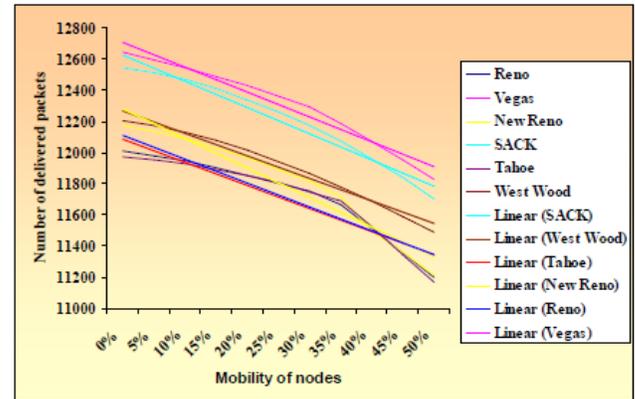

Figure 3: Throughput on different mobility rates at the maximum speed of 5 m/sec with RW

TCP Reno and Tahoe avoid time outs in case of multiple consecutive losses occur. The major factor of degrading the performance for TCP Tahoe has no support of fast recovery algorithm. This algorithm causes to recover the lost segments frequently. Figure 4 shows the throughput performance of TCP Variants at the maximum speed of 10 m/sec. the performance is affected by increasing the speed and TCP New Reno is severely affected, the reason of weak performance for TCP New Reno is also aggressive behavior of fast retransmission algorithm whenever duplicate acknowledgement (dupack) are received and high mobility of nodes is available. Due to aggressive behavior of fast retransmission algorithm, it is difficult to deliver the packets even partial Acknowledgements (Acks) are received to sender. Multiple losses due to high mobility make the weak performance of TCP New Reno because multiple losses cannot be handled properly and network becomes more congested and packets start to drop quickly and this claim is already verified in [4]. Another performance degrading factor relates to TCP New Reno is to take one RTT to perceive each packet loss.

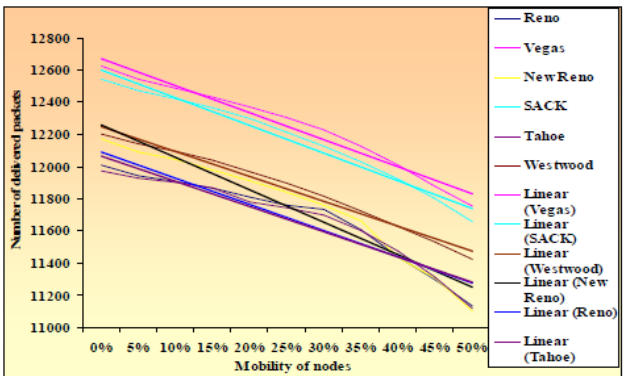

Figure 4: Throughput on different mobility rates at the maximum speed of 10 m/sec with RW

**B. (RTT FAIRNESS OF TCP VARIANTS)**

We show fairness by sharing the bandwidth among different TCP variant according Round trip time (RTT) given in figure 5. There are numerous reasons for RTT fairness as one reason is to attain the equal bandwidth allocation where the different competing flows may allocate similar bottleneck. Long RTT consumes more resources than short RTT in consequence long RTT produces discouraging throughput. TCP Vegas gets higher throughputs than other TCP Variants because slow start and congestion recovery algorithms mostly influence the throughput [12].

Hence slow start and congestion recovery mechanism work in different way for each TCP Variants because TCP Vegas depends on difference of expected and actual throughput. The multiple losses can be retained by avoiding timeouts because the TCP Vegas retransmit the lost segments after receiving 2 (dupack), as that is reason before timeouts, dropped segments are retransmitted and better throughput is obtained. Original feature for TCP Vegas is its congestion detection mechanism because it shows the problems concerning to fairness.

In congestion avoidance, the congestion detection algorithm of TCP Vegas verifies every RTT that is benefit of TCP Vegas over rest of TCP Variants. Moreover TCP Tahoe, TCP Reno, TCP New Reno and SACK reduce the congestion windows more than once during the single RTT that is also reason for unfairness and producing minimum throughputs where as RTT of Vegas reduces only once during the RTT.

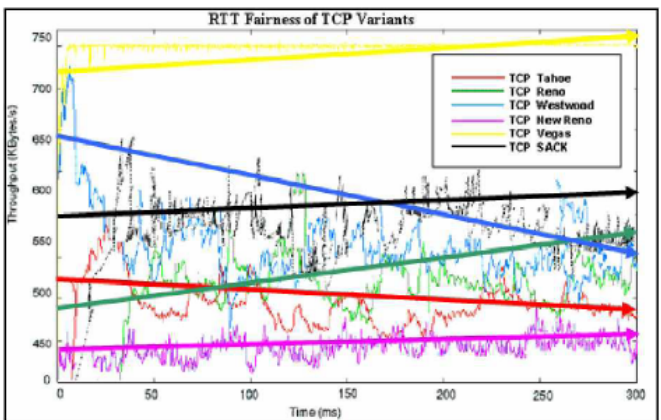

Figure 5: RTT Fairness for TCP Variants at the speed of maximum 10 m/sec with RW

**C. (ENE TO-END DELAY FOR EACH TCP VARIANTS AT SPEED OF 5 & 10 m/Sec).**

End-to-End delay is an average elapsed time for delivery of individual data packets. All possible delays are included and caused by routing discovery, transmission at the MAC layer and queuing at the interface queue, etc but successfully deliveredpackets are calculated. We show trend for each TCP Variants in figure 6 and 7. Vegas has minimum Endto- End delay at the speed of 5m/sec and 10 m/sec whereas TCP Reno and TCP Tahoe have almost similar maximum End-to-End delay at 5 m/sec. As at the speed of 10 m/sec, maximum delay has been analyzed for TCP New Reno. The reason for maximum End-to-End delay for TCP Reno and TCP Tahoe at the speed of 5 m/sec is weakness of faster transmission algorithms. Since TCP Tahoe does not send instant (ACKs) and depends on commutative (ACKs). Therefore when packet is lost then it waits for timeout or pipeline is emptied. This causes high bandwidth delay. TCP Reno behaves like TCP Tahoe whenever multiple losses occur and multiple losses are perceived as single segment losses. Another problem occurs with TCP Reno when the size of window is small; numbers of duplicates (ACKs) are not detected for fast retransmission and have to stay for coarse grained timeout. From other side, TCP New Reno performs weak by increasing the mobility and takes long End-to-End delay. Reason is Limitation of retransmitting single lost segment against per RTT, consequences large delay occurs in retransmitting the later lost packets in the window.

From other side, if the sender is restricted by the receiver's advertised window during recovery time, then the sender is unable to utilize the existing bandwidth successfully and takes long End-to-End delay [14]. Minimum End-to-End delay for TCP Vegas is fairness of retransmission algorithm when segment is lost that TCP Vegas waits for 2 dupack and retransmit the lost segments before expiry of timeouts and RTT of Vegas shrinks only once during the RTT.

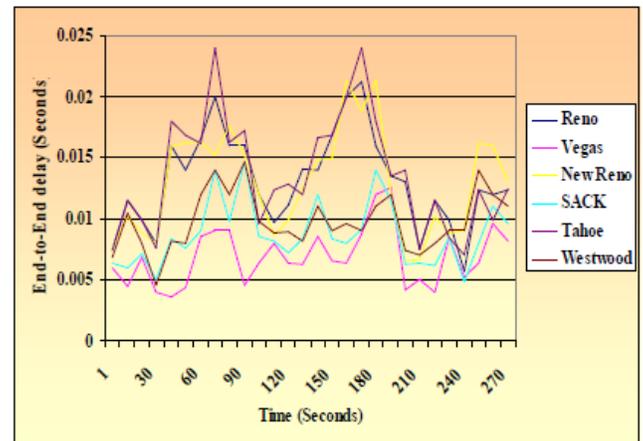

Figure 6: Average End-to-End delay for each TCP Variants at the maximum speed of 5 m/sec with RW Model

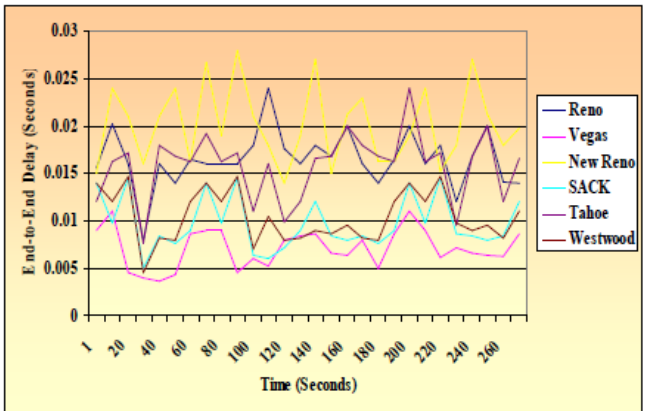

Figure 7: Average End-to-End delay for each TCP Variants at the maximum speed of 10 m/sec with RW Model

### D. (CONTROL OVERHEAD AT DIFFERENT SPEEDS WITH RW MODEL)

"This is the ratio between the total numbers of control packets generated to the total number of data packets received during the simulation time [16]. Control overhead contains control packets, which are used to set up a path to the destination, maintaining and repairing the routes. Control packets are Route Request (RREQ), Route Response (RREP) and Route Error (RERR). Figure 8 shows the trend of control overhead at different mobility ratio. It is very hard to discover a functional route to destination when speed increases. Contention and congestion due to the overflowing behavior of ZRP protocol dominate the effect of the speed. Whenever speed increases that extra routes are needed in ZRP. The overhead of control packets increases significantly as speed increases. Hence more route request segments and route error are transmitted at the higher. Whenever mobility ratio and speed increase that more links break, resulting many control packets are required for route discovery. Due to increasing of mobility ratio and speed that more segments travel over non-optimal routes with larger hop counts, which may be accumulated in a route cache. As a result, these segments will experience longer End-to-End delay and causes the creation of many overhead (control packets [13]. ZRP also creates control overhead packets because it often uses corroded routes due to the large route cache, which causes frequent segments retransmission and very high delay times. ZRP is appropriate for networks in which mobile nodes travel at reasonable speed but not higher. If speed of nodes is increased, resulting more control overhead (control packets) is produced. The behavior of routing protocol, increase in mobility ratio and speed of mobile nodes are three factors, which creates more control overhead (control packets).

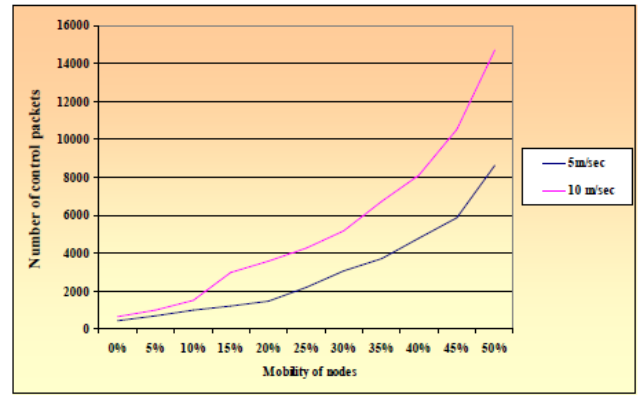

Figure 8: Control Overhead (Number of control Packets) at different speeds with RW Model

### E. (Number of Broken links due to different mobility rates)

We show an average broken links for each TCP Variants at the speed of 5 m/sec and 10 m/sec given in Figure 9. These broken links are calculated at the 50% mobility. When MANET nodes want to establish the sessions to obtain internet services from wired segment of network then routing protocols start route discovery process. Route Request packet (RREQ) is broadcasted into network to obtain any single appropriate route to destination. When route request packet is reached to destination, in response route reply packet (RREP) is sent to originator RREQ. If a link is broken due to mobility and speed of middle nodes, a route error packet is sent to the destination. Meanwhile destination finds another route. The process is repeated until the reply reaches the target. Therefore, destination finds another route if any error occurs in current route. This process causes delay in packet delivery. The high mobility and speed makes more broken links due to discovery of route. The high mobility and speed continuously change the direction of node, inconsequence more links break. Due to increase of speed, topology changes rapidly and more links are broken particularity in ZRP when more connections are established between the nodes.

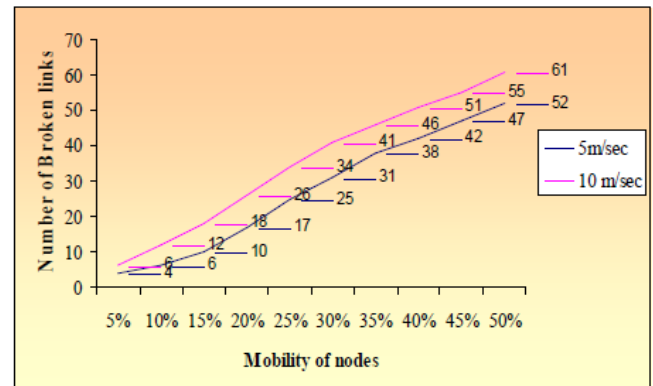

Figure 9: Broken links at different speeds with RW Model

**VIII. DISCUSSION OF RESULTS**

We have simulated TCP Variants in ns-2 over Hybrid network, which consists of MANET, Wired & Wireless segments of network. We have increased mobility in MANET segment of network by using ZRP routing protocol and analyzed that by increasing the mobility and speed the performance of each TCP Variants gradually decreases. Multiple routes obviously give benefits but creates disadvantage due to high mobility. In larger networks, the source routing principle can also generate a trouble. It has been observed that TCP Vegas performed better than other TCP Variants. It produces healthy throughput, better in-order delivery of data, minimum End-to-End delay, and good RTT at different mobility ratios and speeds. The major reasons for degraded performance of TCP Reno, Tahoe and New Reno are timeouts, as during the timeouts period, Variants cannot resend the lost segments whereas TCP Vegas does not wait for loss to trigger cwnd decrease and calculate approximately the current throughput during each [15]. Westwood greatly miscalculates the existing bandwidth, which is potentially troublesome for fairness and can lead to starvation of simultaneous connections that is the reason to produce lesser throughput than TCP Vegas and presented the work in the paper Performance evaluation of TCP Westwood+ [17]. Hence Losses in wired network are due to overflow of buffer at routers. TCP Reno, Tahoe and New Reno have been designed particularly for wired network and meet the requirement of IEEE802.11 but their performance become weak in Hybrid network especially satellite link is involved. Minimum mobility ratios create less control overhead, which causes the better performance for each TCP Variants, which is also proved in our simulation. TCP experiences most losses in multi hop wireless networks, which are caused by packet drop at wireless link layer IEEE 802.11. To improve the performance, new congestion control flavors have been introduced and various schemes are included. Explicit congestion Notification (ECN) has been incorporated to improve the congestion control. If congestion occurs in network then that intermediate routers will mark the congestion experience (CE) code point in header of TCP. This message informs the end host that network is congested and resulting unnecessary packet drops can be prevented.

**IX. CONCLUSION AND FUTURE WORK**

Mobile Adhoc networks (MANET) can be deployed to many locations without the use of infrastructure support. In military environment, disaster situation, scattered educational institutions need such networks to route data packets through dynamically mobile nodes. MANET is better choice for these extremely mobile and dynamic applications, which are not supported by centralized administration. If internet services are required that MANET is better solution in anywhere to integrate with wired network to construct as Hybrid network in order to obtain an internet facility.

To investigate the performance of different transmission control protocols, we have done simulation in ns-2 by using Random Waypoint Mobility model and analyzed different metrics. We have particularly focused on MANET & wired portions of network to investigate the performance of TCP Variants. The minimum effect of mobility has been analyzed on TCP Westwood and reasons are already discussed in detail but it delivers lesser segments than TCP Vegas and SACK whereas TCP Vegas has better throughput, minimum End-to-End delay, better In-order delivery of data and improved RTT. TCP SACK also performs better and does not loss many segments because sender is informed which segment has been received. TCP SACK uses SACK blocks at receiver side to indicate the contiguous block of data successfully received. The sender can find out through SACK blocks which segments are lost, as this is the reason to control the loss of segments frequently. TCP Reno, TCP New Reno and Tahoe degrade the throughput in high mobility ratios and take more End-to-End delay time as compare to other TCP variants and reasons are already illustrated. In future, we will analyze and evaluate TCP Variants in Hybrid network with respect to different mobility models including social network model, Random Walk Mobility Model, Random Direction Mobility Model, City Section Mobility Model etc. We would study under utilized and congested network conditions by using maximizing traffic flows. We would also analyze multihoming issues in future. Finally we suggest if the features of TCP Vegas and TCP Westwood are combined that new Variant could be better from mobility point of view in MANET and mixed environments.

**REFERENCES**

[1] Razaque. A, Shahzad. K, Qadir. M.A, "Performance Evaluation of TCP Variants in mobility based Anchor point node Hybrid Network", IEEE, ACM & SIGWEB, CNSR2008. *In Proceedings of* Sixth Annual Conference on Communication Networks and Services Research (CNSR2008) *IEEE*,ACM & SIGWEB, Halifax, Nova Scotia, CANADA, pp- ,270-277, May 5 - 8, 2008.

[2] Fernando Tapia, "TCP over mobile ad hoc networks", thesis-2004.

[3] Razaque.A,Qadir.M.A, Shahzad.K, Osama Bin Nazeer, Hammad Khalid, "Device Discovery with qualitative and quantitative Analysis of TCP Variants under mobility Based Anchor Point Node Hybrid Network", In Proceedings of IADIS International Conference Wireless Applications and Computing, Amsterdam**,** Netherlands**.** 22 - 24 July 2008.

[4] Qiang Ni,Thierry Turletti and Wei Fu," Simulation-based Analysis of TCP Behavior over Hybrid Wireless and Wired Networks", Project RNRT-VTHD++ , France, October 2002.

[5] A.O. Oluwatope., A. B. Obabire., and G. A. Aderounmet, "End-to-End Performance Evaluation of Selected TCP Variants across a Hybrid Wireless Network", In Information Universe Transactions on Issues in Informing Science and Information Technology, Vol. 3, pp 479-487, 2006.

[6] Camp, T., Boleng, J., Davies, V,A survey of mobility models for ad hoc network research. Wireless Communication and Mobile Computing, Special Issue on Mobile Ad Hoc Networking: Research, Trends and Applications, 2(5): 438-502,2002.

[7] Musolesi, M., Hailes, S., Mascolo, C," An Ad-Hoc Mobility Model Founded on Social Network Theory", Proceedings of the 7th ACM International Symposium on Modeling, Analysis and Simulation of Wireless and Mobile Systems (ACM MSWiM'04). Venice, Italy, p.20-24,2004.

[8] Anthony Lo, Jinglong Zhou, Martin Jacobsson and Ignas Niemegeers, "*ns-2* Models for Simulating a Novel Beyond 3G Cellular Multihop Network", ACM International Conference Proceeding Series; Vol. 202, Article No. 7, Pisa, Italy, October 10, 2006.

[9] Bai, F., Sadagopan, N., Helmy, A., 2003," Important: A Framework to Systematically Analyze the Impact of Mobility on Performance of Routing Protocols for Ad Hoc Networks", Proceedings of the 22nd Annual Joint Conference of the IEEE Computer and Communications Societies (INFOCOM'03), 2:825-835.

[10] Tracy Camp, Jeff Boleng, Brad Williams, Lucas Wilcox, William Navidi," Performance Comparison of Two Location Based Routing Protocols for Ad Hoc Networks", Colorado School of Mines, NSF Grants ANI-9996156 and ANI- 0073699,2003.

[11] Vassilis Tsaoussidis and Ibrahim Matta, "Open Issues on TCP for Mobile Computing", NSF grant Career ANI-0096045, USA, 2002.

[12] Sangtae Ha, Yusung Kim, Long Le, Injong Rhee Lisong Xu," A Step toward Realistic Performance Evaluation of High-Speed TCP Variants", North Carolina State University, 2006.

[13]David Oliver jorg, Torsten Braun," Performance Comparison Of MANET Routing Protocols In Different Network Sizes", computer Science Project, University of Berne ,Switzerland, 2003.

[14]Kevin Fall and Sally Floyd," Simulation-based Comparisons of Tahoe, Reno, and SACK TCP", U.S. Department of Energy under Contract No. DE-AC03-76SF0009, 1996.

[15] Hassan Mahbub, Jain Raj,"High Performance TCP/IP Networking Concepts, Issues, and Solutions (Book)", chapter 11, pp-264-279, 2003.

[14] Dongkyun Kim, Hanseok Bae, Jeomki Song, "Analysis of the Interaction between TCP Variants and Routing Protocols in MANETs", p1, 2005

[15]. Dongkyun Kim, Juan-Carlos Cano , P. Manzoni , and C-K. Toh ," A comparison of the performance of TCP-Reno and TCP-Vegas over MANETs", pp1-2.

[16] O. J. Oyedapo and D. Ngwenya, "Evaluation of TCP-Variants Performances in an Ad-Hoc Mobile Network", pp1,2007.